\documentclass{ifacconf}

\usepackage{graphicx}      
\usepackage{natbib}        
\usepackage{amsmath,amssymb}
\counterwithin*{section}{part}
\usepackage[svgnames,x11names,table]{xcolor}

\makeatletter
\let\old@ssect\@ssect 
\makeatother
\usepackage{natbib}
\usepackage[colorlinks,linkcolor=DarkBlue,urlcolor=DarkBlue,citecolor=DarkBlue]{hyperref}
\makeatletter
\def\@ssect#1#2#3#4#5#6{%
  \NR@gettitle{#6}
  \old@ssect{#1}{#2}{#3}{#4}{#5}{#6}
}
\makeatother

\usepackage{siunitx}
\usepackage{scalerel}
\usepackage{xcolor}
\usepackage{dsfont}
\usepackage{empheq}
\usepackage{bm}
\usepackage{boldmath}
\usepackage{url}
\usepackage{listings}
\usepackage{algorithm}
\usepackage[noend]{algpseudocode}
\usepackage{algorithmicx}
\usepackage{etoolbox} 
\usepackage{graphicx}      
\usepackage{cleveref}
\usepackage{natbib}        
\usepackage{ marvosym }
\usepackage{subcaption}

\begin{document}
\begin{frontmatter}

\title{Sparse State Feedback Control for Industrial Applications\thanksref{footnoteinfo}} 

\thanks[footnoteinfo]{This work was partially supported by the Wallenberg AI, Autonomous Systems and Software Program (WASP) funded by the Knut and Alice Wallenberg Foundation. Authors from the department of automatic control, Lund university, are members of the ELLIIT Strategic Research Area.}

\author[First]{A. Gurpegui Ramón}
\author[First,Second]{F. Norlund} 
\author[First]{K. Soltesz}
\author[First]{A. Rantzer}

\address[First]{Lund University, Dept. Automatic Control, Lund, Sweden 
(email: \{alba.gurpegui\_ramon, frida.norlund, kristian.soltesz, anders.rantzer\}@control.lth.se)}
\address[Second]{Boliden AB, Boliden, Sweden }

\begin{abstract}                
We present an optimization-based methodology for designing sparse state-feedback controllers for industrial applications that are suited for linear control, and demonstrate the framework by designing a level controller for an industrial rougher flotation bank at the Aitik mine. In contrast to the dense linear-quadratic (LQ) controller gains currently operating at the concentrator, our approach enforces a sparsity pattern that is consistent with the interaction structure of the flotation bank and accounts for the worst-case expected inflow disturbances during tuning, while optimizing controller performance through the Integral Absolute Error (IAE) index. The non-zero elements of the sparse gain matrices are optimized using a coordinate search algorithm that handles bound constraints and preserves closed-loop stability. The resulting sparse controller achieves improved load disturbance rejection in the flotation cells compared to the LQ controller. These improvements are consistently observed in both linear and nonlinear simulations. In addition, the imposed structure, results in gain matrices that are easier to adjust and interpret. Importantly, the sparse controllers generated for the Aitik mine are directly suitable for industrial deployment and offer an effective alternative to the existing dense LQ design. 
\end{abstract}


\end{frontmatter}
\section{INTRODUCTION}
\label{sec:introduction}
Classical optimal regulators such as linear-quadratic (LQ) controllers produce dense gain matrices in which every actuator depends on almost every measurement, but in many industrial systems, this level of interconnection is neither necessary nor practical. 

Sparse control formulations address this by enforcing zero-entries in the feedback matrices, restricting interactions to those that are physically motivated or practically feasible. A substantial body of research demonstrates that introducing sparsity into the feedback gain structure can preserve most of the closed-loop performance while simplifying the communication structure (\cite{sparse_feedback}), yielding controllers that are easier to interpret and tune. Although sparse controller synthesis has been extensively studied in the control literature, typically via $H_2$ formulations (\cite{sparse_feedback2, sparse_feedback3}), other performance metrics, such as the integral absolute error (IAE) are standard objectives for controller tuning and benchmarking in process control (\cite{Guzman}).

The present work bridges this gap by addressing sparse controller synthesis in a process-control setting. The non-zero entries of the sparse feedback matrix are tuned through numerical optimization to minimize the IAE and the sparsity pattern is chosen to be consistent with the dynamics of the process to enhance interpretability. This renders implementation-ready sparse state-feedback matrices tailored to the process at hand.

To demonstrate the methodology, we design a sparse state-feedback level controller for a flotation bank in the mining industry. Flotation is a commonly used process to separate valuable minerals from waste rock. In tank cells filled with a slurry of finely ground ore and water, differences in surface properties are utilized to separate the different minerals. The selected minerals float to the top of the cell, assisted by air bubbles, forming a mineral froth on top of the cell that is collected as it flows over the rim. Hence, good level control is a foundation for receiving good mineral recovery. 

Flotation cells rarely operate as stand-alone units, but rather in series of cells to receive maximum mineral recovery. This setup makes a flotation bank an interconnected system where different process variables affect each other through interaction. 
A survey from \cite{LEROUX202011854} showed that around 60~\% of industrial flotation banks are today controlled by multi-input-multi-output (MIMO) structures. Although the sample size is small and skewed towards South Africa, this result indicates a general trend in the industry where traditional control structures, consisting of  proportional-integral-derivative (PID) controllers, give way to MIMO controllers such as LQ controllers or model-predictive-controllers (MPC). 

One example of a successful change of controllers, was reported by \cite{Norlund250703} where an LQ-controller successfully replaced a PI-control structure for the level control of the rougher flotation bank in Boliden AB's concentrator plant at the Aitik mine in late 2023. The LQ controller has been the default controller in the plant since then, with an up-time of over 96~\%.

During the commissioning phase of the LQ controller at Aitik, questions were raised, both from engineers and operators, of whether the matrices needed to be dense as the dense nature of the matrices makes fine tuning of the controller by the operators hard. Hence, the day-to-day controller maintenance and fine tuning traditionally performed by plant personnel becomes impractical, if at all possible, making it harder for plant personnel to take ownership for the new controller and its maintenance.

\section{PROCESS DESCRIPTION} 
\label{sec:exampleProcess}

\begin{figure}
    \centering
    \includegraphics[width=\linewidth]{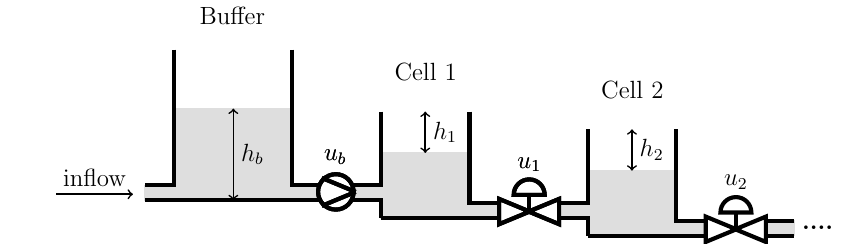}
    \caption{A process schematic of the flotation process.}
    \label{fig:process_description}
\end{figure}

The first flotation step in the Aitik concentration plant, the rougher flotation, consists of two identical flotation banks, each containing a buffer tank and 13 consecutive identical flotation cells. A schematic sketch of the process can be seen in \cref{fig:process_description}, where the buffer tank and the first two flotation cells are shown.

The slurry that enters the flotation process from the two upstream milling lines is divided equally between the two flotation lines. Hence, whenever a milling line experience disturbances, the resulting flow disturbance affects both flotation lines. The cancellation of these inflow disturbances was one of the main motivators for the control structure change described in \cite{Norlund250703}. 

\begin{figure}
    \centering
    \includegraphics[width=0.9\linewidth]{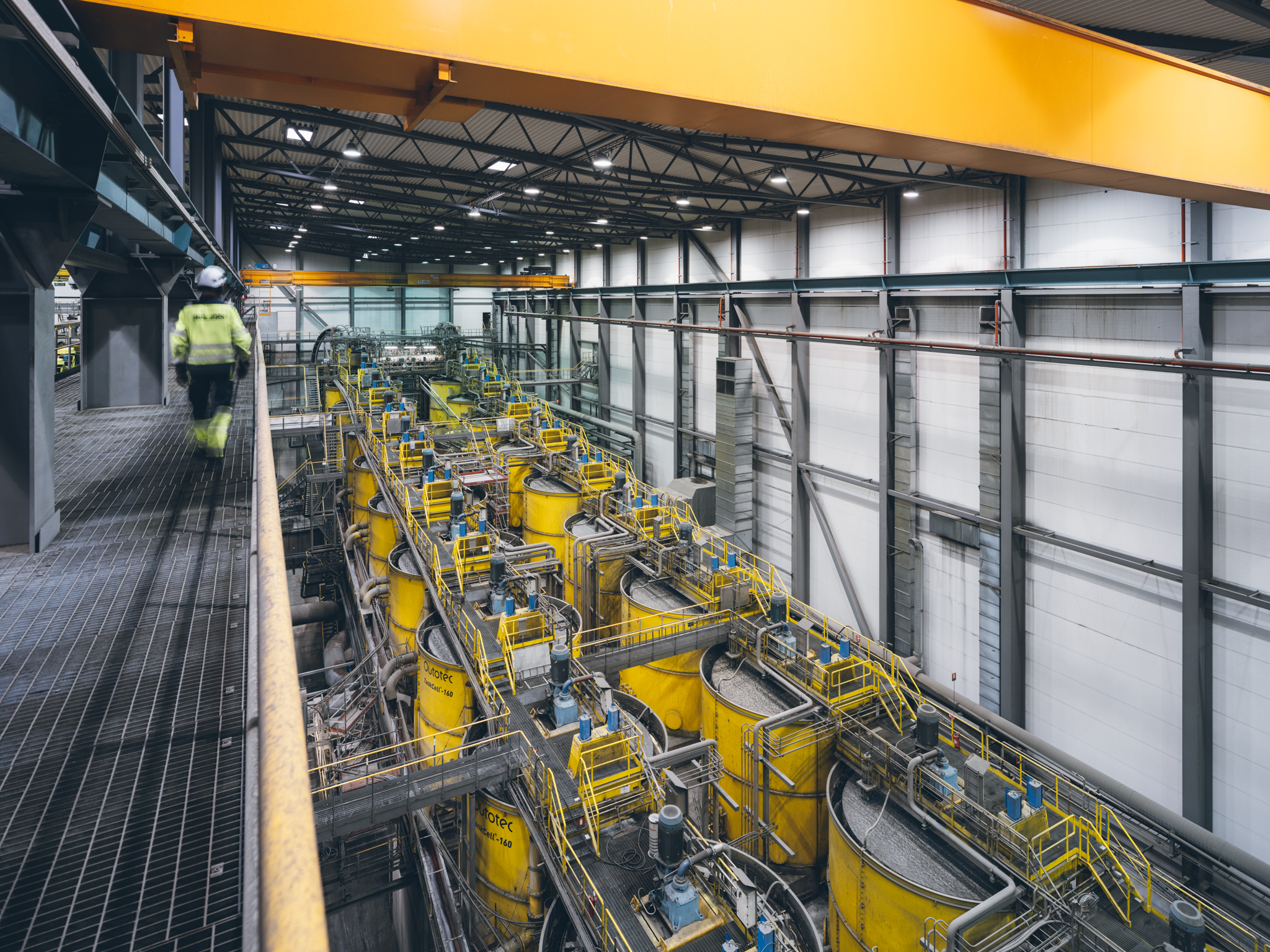}
    \caption{The two parallel rougher flotation lines at the Aitik concentrator plant. Photo: Jonas Westling}
    \label{fig:Aitik}
\end{figure}

The inflow from the milling lines to each flotation bank is continuous, but not measurable, and it is hence treated as an unknown load disturbance. From the buffer tank, the slurry is pumped to the first flotation cell. The 13 flotation cells are identical, and separated by valves. The flow through the valves is driven by the physical level difference of one meter between consecutive cells. A picture from the plant can be seen in \cref{fig:Aitik}.

The level $h_i$ in tank $i$ in the process can be modeled as
\begin{equation}
    \dot h_i = \frac{1}{A_{tank}}(h_i-h_{i+1}+1)f(u_i),
\end{equation}
where $A_{tank}$ is the cross section area of the tank, $f(u_i)$ is a function that relates the control signal of the outlet valve of the tank to the flow through it. For the flotation cells, this function is approximately linear, while for the buffer tank pump, the function is quadratic. It is also notable that all units in the model are true to the units of the Aitik concentration plant.
\section{METHODOLOGY}
\label{sec:theory}

\subsection{Linear plant model}
\label{subsec:plant_model}
The flotation process is meant to be operated at a steady operating point, and hence the nonlinear process model described in \cref{sec:exampleProcess} is linearized around a nominal point $(\Bh_0,\Bu_0)$. The resulting linearized system is 
\begin{equation}
\label{eq:lin_model}
    \dot{\Bx}(t) = A \Bx(t) + B \Bu(t),
\end{equation}
where $\Bx \in \mathbb{R}^{14}$ represents the deviations of the buffer tank and the flotation cell levels from the steady state, and $\Bu \in \mathbb{R}^{14}$ denotes the deviation in control inputs to the pump and the valves. The system matrices $A$ and $B$ are given by
\begin{align*}
    A =
\begin{bmatrix}
0 & 0 & 0 & \cdots & 0 \\
0 & \alpha & \delta & \cdots & 0 \\
0 & \epsilon & \beta & \delta &\\
\vdots & & \ddots & \ddots & \ddots \\
0 & 0 & \cdots & \epsilon & \beta
\end{bmatrix},\qquad
B =
\begin{bmatrix}
\zeta_{\mathrm{bt}} & 0 & 0 & \cdots & 0 \\
\lambda & \eta & 0 & \cdots & 0 \\
0 & \mu & \eta & \ddots & \vdots \\
\vdots & & \ddots & \ddots & 0 \\
0 & 0 & \cdots & \mu & \eta
\end{bmatrix}
\end{align*}
with 
\begin{align*}
\alpha &= -0.0147, &\beta &= -0.0295, &\delta &= 0.0147,  \\
\epsilon &= 0.0147, &\lambda &= 0.00934, &\eta &= -0.0491,  \\
 \mu &= 0.0491, &\zeta_{\mathrm{bt}} &= -6.37\cdot10^{-4}.
\end{align*}
This results in a state-space representation that is sparse and banded as each flotation cell interacts only with its immediate neighbors. As the flow to the first flotation cell is pumped from the buffer tank, the dynamical model of this interaction differs from the interactions between the downstream flotation cells. 
To eliminate steady-state level offsets, we augment the model with integral states, 
\begin{equation}
    \Bz(t) = \int_0^t (\Br - \Bx(\tau))\,d\tau.
\end{equation}
Choosing the linearization point such that $\Br = \mathbf 0$, we have 
\begin{equation}
    \Bz(t) = -\int_0^t \Bx(\tau)\,d\tau,
\end{equation}
and each component $z_i(t)$ represents the accumulated deviation of the level $i$ from its nominal value. Note that the augmented state space representation is described by
\begin{equation}
\label{eq:x_aug}
    \dot{\Bx}_{\mathrm{aug}}(t) =
    \underbrace{\begin{bmatrix}
        A & 0 \\[1mm]
        -I_{14} & 0
    \end{bmatrix}}_{A_{\mathrm{aug}}}
    \Bx_{\mathrm{aug}}(t)
    +
    \underbrace{\begin{bmatrix}
        B \\[1mm]
        0
    \end{bmatrix}}_{B_{\mathrm{aug}}}
    \Bu(t),
    \quad
    \Bx_{\mathrm{aug}}(t) = \begin{bmatrix} \Bx(t) \\ \Bz(t) \end{bmatrix}.
\end{equation}
\subsection{Existing LQ controller}
With the linear model of the plant described in \cref{subsec:plant_model}, an LQ controller with integral action can be designed.
The infinite-horizon quadratic cost is
\begin{equation}
\label{eq:LQ_cost}
    J_{\mathrm{LQ}} =
\int_0^\infty \Big(
\Bx^\top Q_x \Bx + \Bz^\top Q_z \Bz + \Bu^\top R\,\Bu
\Big) \, dt.
\end{equation}
The controller running in the Aitik plant minimizes \cref{eq:LQ_cost} for 
\begin{align*}
&\scalebox{0.95}{$Q_x =
\mathrm{diag}\!\begin{bmatrix}
1.0 &
0.0123 &
\cdots &
0.0123
\end{bmatrix}$}\\
&\scalebox{0.95}{$Q_z =
\mathrm{diag}\!\begin{bmatrix}
6.86\cdot10^{-6} &
1.29\cdot10^{-5} &
1.08\cdot10^{-5} &
\cdots &
1.08\cdot10^{-5}
\end{bmatrix}$},\\
&\scalebox{.95}{$R =
\mathrm{diag}\!\begin{bmatrix}
0.04 &
0.0204 &
\cdots &
0.0204
\end{bmatrix}$}.
\end{align*}
These weights were chosen as they provide an adequate balance between the buffer tank deviation and the inflow disturbance in the flotation cells. Note that all flotation cells are treated equally in the cost function as they are identical. For more details on the LQ tuning process, see \cite{Norlund250703}.

Minimizing \cref{eq:LQ_cost} subject to the state dynamics in \cref{eq:x_aug} gives an optimal LQ controller given by
\begin{align*}
\Bu(t) = -K_{\mathrm{d}} \Bx(t) + K_{I,\mathrm{d}} \Bz(t),
\quad
[K_{\mathrm{d}} \;\; -K_{I,\mathrm{d}}] = R^{-1} B_{PI}^\top P,
\end{align*}
where $P$ solves the algebraic Riccati equation
\begin{align*}
0&=A_{aug}^\top P + P A_{aug} - P B_{aug} R^{-1} B_{aug}^\top P + Q_{aug} 
\end{align*}
with $Q_{aug}=\mathrm{blockdiag}(Q_x,Q_z)$. 

\subsection{Sparse Controller Parametrization}
\label{subsec: sparse controller}
The feedback gain matrices $K_{\mathrm{d}}$ and $K_{I,\mathrm{d}}$, resulting from the LQ controller design, are dense, and while they show good performance, the dense structure is undesirable for interpretation and maintenance. Therefore, we aim to implement a sparse design parametrization. To do this, we impose a structure on the gain matrices through a mask that only allows nonzero elements on the three central diagonals and in the first column. I.e.,
\begin{align*}
    [K_s]_{ij} = [K_{I,s}]_{ij} = 0
\quad \text{whenever} \quad j \notin \{0, i-1, i, i+1\}.
\end{align*}
This structure is process specific, the main diagonal governs the self-feedback of each controlled variable, the sub-diagonal serves as a feedforward connection to the cell one position upstream, while the super-diagonal accounts for disturbances passed on to the next downstream cell. The first column coordinates the inflow disturbance response of the flotation bank. The structure of the non-zero elements in the masks are chosen based on engineering knowledge.

As the inflow to the buffer tank is subject to disturbances when the milling lines don't run smoothly, a model of the known worst-case disturbance acting through the buffer tank, together with its resulting propagation throughout the flotation cells, is explicitly incorporated into the sparse controller synthesis.
\begin{align*}
\dot{\Bx}(t) = A \Bx(t) + B \Bu(t) + H g(t),
\end{align*}
with $H = \xi \cdot \mathds{1},\quad \xi = 2.6\cdot10^{-7}$ and a piecewise-constant disturbance input
\begin{align}
\label{eq:g(t)}
g(t) =
\begin{cases}
 -5.5\cdot 10^5, & 3000 \le t \si{[s]} \le 7000, \\[1mm]
 0, & \text{otherwise}.
\end{cases}
\end{align}
This corresponds to when one milling line comes to an abrupt complete stop, resulting in a reduction in the inflow by $50\%$. The somewhat peculiar unit of $g(t)$, \si{cm^3}, is a consequence of the measurements in the process being given in \si{cm}.
Introducing this disturbance does not change the structure of the sparse controller, which is given by
\begin{align*}
    \Bu(t) &= -K_{\mathrm{s}}(p) \Bx(t) + K_{I,\mathrm{s}}(p) \Bz(t), \\
\dot{\Bz}(t) &= -\Bx(t).
\end{align*}
To better reflect the desired disturbance rejection properties, we tune the sparse controller in a simulation scenario with the inflow disturbance described in \cref{eq:g(t)}, and use a weighted IAE objective,
\begin{align*}
J_{\mathrm{IAE}}(p) = \int_0^{T} \|Q_{\mathrm{IAE}} \hspace{0.5mm}\Bx(t)\|_1\,dt,
\end{align*}
where $Q_{\mathrm{IAE}} = \mathrm{diag}(\alpha, 1,\dots,1)$ and $\alpha = 1/30$.
The first diagonal entry weights the buffer tank, while the remaining entries weight the flotation cells. Having a different penalty factor $\alpha$ for the buffer tank allows it to exhibit different behavior compared to the flotation cells. The flotation cells are to be kept close to their references, while the buffer tank is intended to use its volume to buffer, allowing deviations from its reference in favor of smoother inflow changes to the flotation cells. The parameter $\alpha$ is chosen such that the buffer tank exhibits the same behavior in response to inflow disturbances in the IAE optimized framework as it does in the LQ framework. 

To be able to compare the resulting sparse controller to the deployed LQ controller, and to generate a controller that is safe for deployment, the individual gains of the new controller needs to be kept reasonably small, hence the gains need to be limited in the tuning process.

While this could be achieved by incorporating a noise model in the tuning process, it is here achieved by imposing magnitude constraints on the elements in the gain matrices.
We apply element-wise box constraints
\begin{align*}
    |K_{s}| \leq  K_{b}, \qquad
    |K_{I, s}| \leq K_{b},
\end{align*}
where
\begin{align}
\label{eq:bounds ki kis}
    \scalebox{.72}{$K_{\text{b}} =
    \begin{bmatrix}
    8.0  & 0.8 & 0 & 0   & \cdots & 0 \\
    1.6  & 1.0 & 0.4 & 0   & \cdots & 0 \\
    1.6  & 1.0 & 1.0 & 0.4 & \ddots & \vdots \\
    1.6  & 0   & 1.0 & 1.0 & \ddots & 0 \\
    \vdots & \ddots & \ddots & \ddots & \ddots & 0.4 \\
    1.6  & 0   & \cdots & 0 & 1.0 & 1.0
    \end{bmatrix}, \hspace{2mm} K_{I,\text{b}} =
\begin{bmatrix}
0.03  & 0.006 & 0     & 0     & \cdots & 0 \\
0.003 & 0.03  & 0.013 & 0     & \cdots & 0 \\
0.003 & 0.03  & 0.03  & 0.013 & \ddots & \vdots \\
0.003 & 0     & 0.03  & 0.03  & \ddots & 0 \\
\vdots & \ddots & \ddots & \ddots & \ddots & 0.013 \\
0.003 & 0     & \cdots & 0     & 0.03 & 0.03
\end{bmatrix}.$}
\end{align}
The non-zero elements of $K_b$, $K_{I,b}$ come from engineering knowledge at Aitik. These values reflect tuning experience, known actuation limits and acceptable gain ranges. Without these bounds, the optimizer would produce values that are not meaningful or implementable.

To preserve the sign pattern of the dense controller, an auxiliary function is applied before optimization. In this function, the bounds of selected parameters are adjusted so that each sparse coefficient retains the same sign as in the dense gain matrix.

We also impose that the parameters in the gain matrices that represents the same interaction between cells take the same value regardless of the position along the flotation bank, the resulting sparse controller are denoted $(K_{opt}^*, K_{I,opt}^*)$ and will have the structure 
\begin{align}
\label{eq:structure KsKis}
    \scalebox{.8}{$
    K_{opt} =
   \begin{bmatrix}
    b   & f   & 0   & 0   & \cdots & 0 \\
    a   & d   & e   & 0   & \cdots & 0 \\
    a   & c   & d   & e   & \ddots & \vdots \\
    a   & 0   & c   & d   & \ddots & 0 \\
    \vdots & \ddots & \ddots & \ddots & \ddots & e \\
    a   & 0   & \cdots & 0 & c   & d
    \end{bmatrix},
    \hspace{4mm}
    K_{I,opt} =
    \begin{bmatrix}
    l   & k   & 0   & 0   & \cdots & 0 \\
    m   & i   & j   & 0   & \cdots & 0 \\
    g   & h   & i   & j   & \ddots & \vdots \\
    g   & 0   & h   & i   & \ddots & 0 \\
    \vdots & \ddots & \ddots & \ddots & \ddots & j \\
    g   & 0   & \cdots & 0 & h   & i
    \end{bmatrix}.
    $}
\end{align}

Hence, the different parameters to be tuned are
\begin{equation}
\label{eq:p_vector}
    \Bp \;=\; [\,a,\; b,\; c,\; d,\; e,\; f,\; g,\; h,\; i,\; j,\; k,\; l,\; m\,]^\top.
\end{equation}

The sparse tuning problem can thus be formulated as
\begin{align}
\label{eq:IAE_min_formulation}
\min_{\Bp} \quad & J_{\mathrm{IAE}}(\Bp) \notag \\
\text{s.t.} \quad & \Bp_{\min} \le \Bp \le \Bp_{\max},
\end{align}
where $\Bp_{min}$ and $\Bp_{max}$ are obtained by mapping the elements of \cref{eq:bounds ki kis} to vectors in the same way as the elements of \cref{eq:structure KsKis} are mapped to \cref{eq:p_vector}.
We also require that $K_{\mathrm{s}}(\Bp)$, $K_{I,\mathrm{s}}(\Bp)$ yield a stable closed loop.

Equation \cref{eq:IAE_min_formulation} is numerically minimized using a coordinate search~\citep{alg_search}, see Algorithm~\ref{algorithm}. In this method, the initial parameter vector
\begin{align*}
\Bp_0
&= \bigl[a_0,\; b_0,\; c_0,\; d_0,\; e_0,\; f_0,\; g_0,\; h_0,\; i_0,\; j_0,\; k_0,\; l_0,\; m_0\bigr]^\top
\end{align*}
is obtained by projecting the dense gains $(K_{\mathrm{d}}, K_{I,\mathrm{d}})$ onto the sparse structures $(K_{\mathrm{s}}, K_{I,\mathrm{s}})$. Each scalar parameter in $\Bp_0$ is initialized as the average of the dense entries corresponding to its position in the sparse pattern. For instance, $a_0$ corresponds to the average of the off-diagonal elements in the first column of $K_{\mathrm{d}}$.

In the algorithm, each parameter $p_j$ has an adaptive step size $\Delta_j > 0$, 
initialized as $\Delta_j = \max\{0.1|p_j|,\,0.05(p_{\max})_j\}$. This initialization ensures that the local search begins with a perturbation of at least 5\% of the admissible upper bound, or 10\% of the current parameter magnitude, whichever is larger. Consequently, the step along each coordinate is neither too small nor too large. The step size defines a small search interval (trust region) along that coordinate. At every iteration, the algorithm visits all coordinates once in a random order and tests two possible moves, $p_j+\Delta_j$ and $p_j-\Delta_j$, while keeping all other parameters fixed. If a trial value exceeds its limits, it is clipped back to the nearest admissible value. In particular,
\begin{align*}
   \scalebox{.9}{$\mathrm{clip}(q_j,[p_{\min}]_j,[p_{\max}]_j)
   =\begin{cases}
    [p_{\min}]_j, & q_j < [p_{\min}]_j,\\[3pt]
    q_j,          & [p_{\min}]_j \leq q_j \leq [p_{\max}]_j,\\[3pt]
    [p_{\max}]_j, & q_j > [p_{\max}]_j
\end{cases} $}
\end{align*}
is applied separately to each component of $\Bp$. If no improvement is found during a complete sweep, the step sizes are reduced as $\Delta \leftarrow \gamma\Delta$, with $\gamma = 0.7$. Thus, when no improvement occurs in a full pass, we shrink the step size to refine the solution around the current point. The process terminates when all $\Delta_j$ become sufficiently small.
\begin{algorithm}[t]
\caption{Coordinate search}
\begin{algorithmic}[1]
\State $\Bp \gets \mathrm{clip}(\Bp_0,\,\Bp_{\min},\,\Bp_{\max})$
\State $\Delta_j^{(0)} \gets \max\{0.1|p_{0,j}|,\,0.05(p_{\max})_j\}$ for all $j$
\State $\Delta \gets \Delta^{(0)}$
\State $J^\star \gets J(\Bp)$
\For{$t = 1,2,\dots,N$}
   \State $\textit{improved} \gets \textbf{false}$
   \State $\pi \gets$ random permutation of $\{1,\dots,n_p\}$
   \For{each $j$ in $\pi$}
      \For{$s \in \{+1,-1\}$}
         \State $\Bp^{\mathrm{cand}} \gets \Bp$
         \State $p^{\mathrm{cand}}_j \gets 
                \mathrm{clip}\bigl(p_j + s\,\Delta_j,\,(p_{\min})_j,(p_{\max})_j\bigr)$
         \State $J^{\mathrm{cand}} \gets J(\Bp^{\mathrm{cand}})$
         \If{$J^{\mathrm{cand}} < J^\star$}
            \State $\Bp \gets \Bp^{\mathrm{cand}}$, \quad $J^\star \gets J^{\mathrm{cand}}$
            \State $\textit{improved} \gets \textbf{true}$
         \EndIf
      \EndFor
   \EndFor
   \If{\textbf{not} \textit{improved}}
      \State $\Delta \gets \gamma\,\Delta$
      \If{$\max_j \Delta_j < 10^{-4} \,\max_j \Delta^{(0)}_j$}
         \State \textbf{break}
      \EndIf
   \EndIf
\EndFor
\State \textbf{return} $\Bp$
\end{algorithmic}
\label{algorithm}
\end{algorithm}
\section{RESULTS}
\label{sec:results}
Applying Algorithm~\ref{algorithm} to the simulation scenario described in \cref{subsec: sparse controller}, i.e. \cref{eq:IAE_min_formulation} is optimized for a 12000 \si{s} long simulation scenario where the system is subject to the disturbance in \cref{eq:g(t)},  we obtain
\begin{align*}
&\Bp_{K_s}^\star =
[\, a,\; b,\; c,\; d,\; e,\; f\,] \\[2mm]
&\;\approx\;
[\, -1.31,\; -6.91,\; -1,\; -1,\; 0,\; 0.8\,], \\[2mm]
&\Bp_{K_{I,s}}^\star =
[\, g,\; h,\; i,\; j,\; k,\; l,\; m\,]\\[2mm]
&\;\approx\;
[\, 2\cdot10^{-3},\; -0.03,\; -0.03,\; 0.01,\; 0.01,\; -0.01,\; 2\cdot10^{-3}\,].
\end{align*}
Note that $e=0$ is set by the algorithm, it is not forced to $0$ by the sparsity mask. 
\begin{table}[]
        \caption{Cross-evaluated closed-loop costs for $x_0=10\cdot \mathds{1}$, $z_0=0$.}
    \centering
    \renewcommand{\arraystretch}{1.2} 
    \setlength{\tabcolsep}{6pt}       
    \begin{tabular}{|r|r|r|r|}
    \hline
       \normalsize Controller/ Cost & \normalsize $J_{IAE}$ & \normalsize $J_{LQ}$ \\
       \hline
        \normalsize LQ -\; $(K_{\mathrm{d}}, K_{I,\mathrm d})$   &  $45.0\cdot10^{3}$ & $21.2\cdot10^6$ \\
        \normalsize Masked LQ -\;$(K_s, K_{I, s})$   &  $146\cdot10^{3}$ & $22.9\cdot10^6$ \\
        \normalsize Sparse -\;$(K_{opt}^*, K_{I, opt}^*)$   & $34.0\cdot10^{3}$ & $20.4\cdot10^6$\\
      \hline
    \end{tabular}
    \label{tab:costs}
\end{table}
Table~\ref{tab:costs} summarizes $J_{IAE}$ and $J_{LQ}$ for the different controllers. The costs are evaluated in the tuning simulation scenario for the sparse feedback controller. Note that the $J_{LQ}$ presented here is evaluated over the finite simulation time of the tuning scenario and the values are hence not to be confused with the classical LQR cost evaluated over an infinite time horizon.

The optimized sparse controller yields a smaller finite-horizon quadratic cost $J_{\mathrm{LQ}}$, approximately 3.8\% lower than that of the dense LQ. At the same time, the sparse controller achieves a 24\% reduction of $J_{IAE}$ compared to the dense LQ. The masked LQ controller overall performs worse, as is expected. These results confirm that direct optimization of the sparse structure is crucial to recover, and in some aspects improve, the nominal performance of the dense controller.
\begin{figure*}[t!]
    \centering
    \begin{subfigure}[t]{0.32\textwidth}
        \centering
        \includegraphics[width=\textwidth]{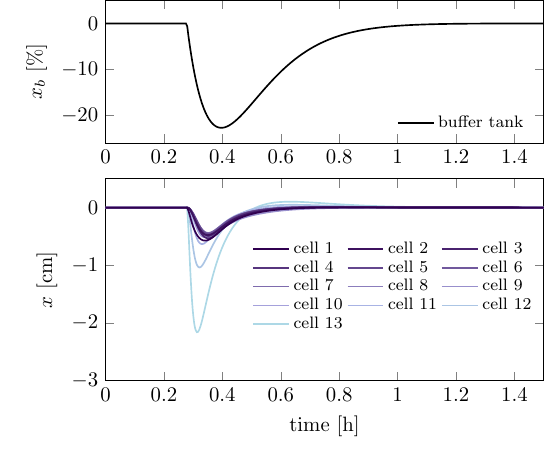}
        \caption{ LQ}
        \label{fig:dist_LQ_Lin}
    \end{subfigure}%
    ~ 
    \begin{subfigure}[t]{0.32\textwidth}
        \centering
        \includegraphics[width=\textwidth]{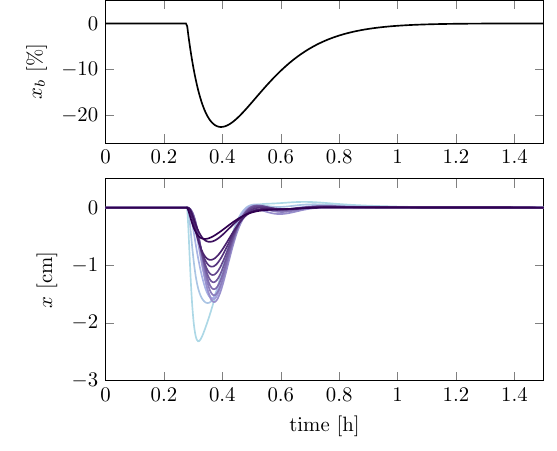}
        \caption{Masked LQ}
        \label{fig:dist_sparse_Lin}
    \end{subfigure}
        ~ 
    \begin{subfigure}[t]{0.32\textwidth}
        \centering
        \includegraphics[width=\textwidth]{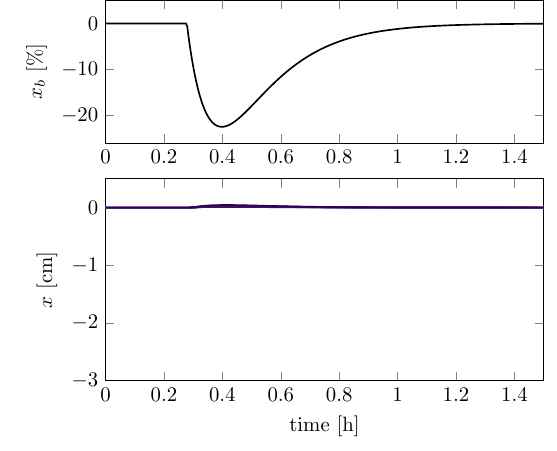}
        \caption{Optimized sparse controller}
        \label{fig:dist_opt_Lin}
    \end{subfigure}
    \caption{The inflow disturbance response of the level of the buffer tank and the levels of the flotation cells. At time 0.3 the inflow to the buffer tank is reduced by 50\%. The simulation model is the linear model. }
    \label{fig:Disturbance_rejection_linear}
\end{figure*}

\begin{figure*}[t!]
    \centering
    \begin{subfigure}[t]{0.32\textwidth}
        \centering
        \includegraphics[width=\textwidth]{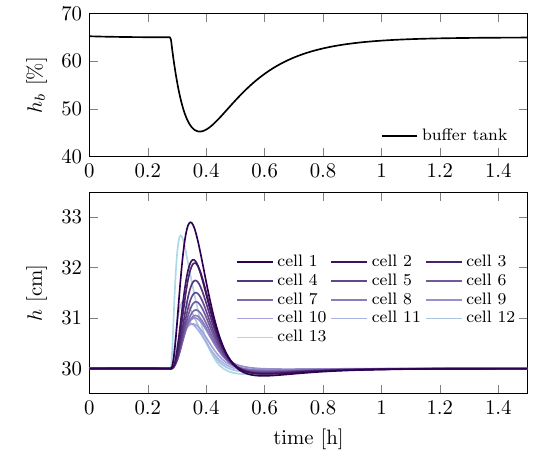}
        \caption{ LQ}
        \label{fig:dist_LQ}
    \end{subfigure}%
    ~ 
    \begin{subfigure}[t]{0.32\textwidth}
        \centering
        \includegraphics[width=\textwidth]{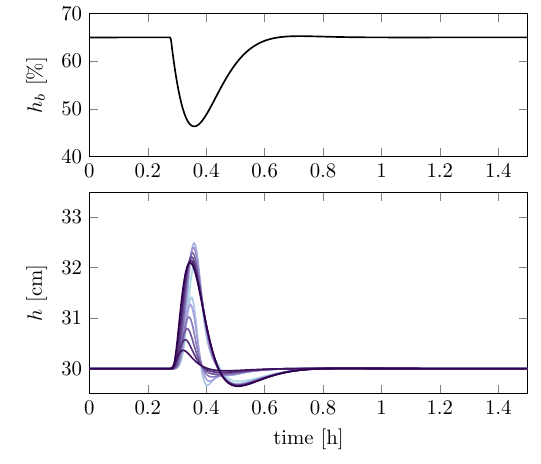}
        \caption{Masked LQ}
        \label{fig:dist_sparse}
    \end{subfigure}
        ~ 
    \begin{subfigure}[t]{0.32\textwidth}
        \centering
        \includegraphics[width=\textwidth]{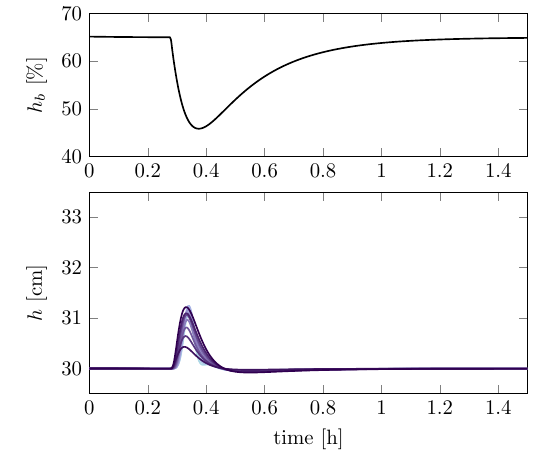}
        \caption{Optimized sparse controller}
        \label{fig:dist_opt}
    \end{subfigure}
    \caption{The inflow disturbance response of the level of the buffer tank and the froth thickness of the flotation cells. At time 0.3 the inflow to the buffer tank is reduced by 50\%. The simulation model is the non-linear model.}
    \label{fig:disturbance_rejection}
\end{figure*}

To further evaluate the disturbance rejecting properties of the controller, \cref{fig:Disturbance_rejection_linear} and \cref{fig:disturbance_rejection} show the isolated response to an inflow disturbance corresponding to a milling line stop. This gives the effect that the inflow to the buffer tank is reduced by 50 \%. In \cref{fig:Disturbance_rejection_linear}, the simulation model is the linear model in \cref{eq:lin_model}, while the simulation results shown in \cref{fig:disturbance_rejection} are generated from the non-linear simulation model described in \cref{sec:exampleProcess}.

The sparse controller is tuned to preserve the buffering behavior of the buffer tank, which is demonstrated by the similar behaviors of the buffer tank for the different controllers in \cref{fig:Disturbance_rejection_linear}. The disturbance rejection for the flotation cells is improved by the optimized sparse controller, as can be seen when comparing \cref{fig:dist_opt_Lin} to \cref{fig:dist_LQ_Lin,fig:dist_sparse_Lin}.

The improved disturbance rejecting properties are also carried over to the non-linear simulation scenario where the disturbance rejecting properties of the sparse optimized controllers remain visibly better than for the LQ controller. This can be seen when comparing \cref{fig:dist_opt} to \cref{fig:dist_LQ,fig:dist_sparse}.
\section{DISCUSSION}
\label{sec:discussion}
As the presented framework aims to design implementation ready controllers tailored to the process at hand, some important factors for applicability are discussed below.

\subsection{Interpretability and Maintenance}
The approach in this paper is presented through an example where the controller is tailored to an example process, and while the resulting controller is case specific, the methodology is applicable beyond this specific process. To adapt to other applications, the sparsity mask need to be tailored to the specific process dynamic to ensure that the remaining values represent meaningful interactions. 

One advantage of the optimized sparse controller is the interpretability of the feedback matrices. Each parameter corresponds to a meaningful interaction in the process and entries related to identical flotation cells hold the same value, hence, the number of parameters to tune is small.

The sparse structure makes fine-tuning of the matrices more accessible, which is essential for the day to day maintenance of the controller. As the sparse structure allows the plant personnel to treat one process variable at a time, necessary adjustments to the controller can be done when process components are exchanged for components with different dynamics. This tuning possibility enables the plant personnel to maintain their ways of working while taking ownership of the new controller. Having local ownership from plant personnel is essential for the controller to last the test of time, as many model based controllers need continuous updates as the physical process changes over the lifetime of the plant.

\subsection{Disturbance Rejection}
While the traditional LQR design provides a well established framework to design controllers, it does not necessary reflect all the desired controller properties in the cost function. In many industrial processes, the disturbance rejecting properties are of great importance, and the IAE is hence well suited as a design criterion in the optimization of the controller parameters. For the example at hand, \cref{fig:Disturbance_rejection_linear} demonstrates that we can achieve better disturbance rejection of unmeasurable inflow disturbances by re-designing the cost function for which we optimize the controller.

Moving from the model-error-free linear simulation to the nonlinear simulation, shown in \cref{fig:disturbance_rejection}, we observe that the improved disturbance rejecting properties are carried over nicely for the flotation process. This is expected since the dynamics of the flotation process are well captured by a linear model. Similar results can be expected for processes that are well represented by linear models.

In \cref{fig:Disturbance_rejection_linear}, the last cells in the flotation series deviate more from the reference as a result of the inflow disturbance than the upstream cells. These end-of-series-effects in the linear simulation are due to that cells in the beginning or middle of the flotation bank take input from neighbors both up- and down-stream through the dense LQ state feedback matrix while the last flotation cells have no, or fewer, downstream neighbors. In the non-linear simulation these effects are reduced and in the real process, where even greater model errors are present, along with process noise, these end-of-series-effects are not noticeable. 

\subsection{Controller Design Choices}
As the controller is intended to be implemented in the real process, the gain values in the matrices needed to be regulated not to take unreasonably high values. This is imposed in the optimization through the bounds in \cref{eq:bounds ki kis}. The bounds were set based on hands on process experience of historical gains. There are two main reasons for imposing such bounds: to keep the reactivity to process noise down, and to keep the rate of change of the control signal in a compatible range with the valves in the real process such that the actuators can match the rate of change of the controller output.

We also forced a preserved sign structure of the sparse feedback matrices compared to the initial LQ controller to preserve the explainable purpose of each entry in the feedback matrices. This is important to gain trust within an organization, and to enable the plant personnel to interact with the controller.

When tuning the controller in the simulation, the worst case disturbance incorporated in the simulation, $g(t)$, enters in all nodes simultaneously. This is a theoretical worst case scenario, where the dampening effects of the buffer tank is by-passed. This does not correspond to the simulation scenario of the milling line stop for which \cref{fig:Disturbance_rejection_linear,fig:disturbance_rejection} are generated. In these cases, the inflow disturbance enters only in the buffer tank, the inflow disturbance that enters the flotation cells is smoothed out by the buffer tank. Hence the design disturbance is worse than the expected worst case disturbance in the real process. This was chosen since designing with a disturbance that enters directly in the flotation cells improves the robustness and the disturbance rejecting properties of the designed controller.

\subsection{Future Work}
The sparse controller is to be tested in the Aitik concentrator plant in the near future. This pilot-implementation will indicate where the framework needs further development. Some identified topics of interest are reference tracking and noise sensitivity. The controller response to reference changes needs to be examined in order to determine if the unmodified reference change response is sufficient or if a feedforward filter for reference changes needs to be added to maintain sufficient performance.
When it comes to noise sensitivity, the current design procedure does not consider a noise model, this could be included to complement the parameter bounds in the design process to address noise sensitivity at the design stage of the sparse controller.

\section{Conclusions}
We have shown that sparser, more interpretable feedback controllers can match, or even improve the control performance with respect to disturbance rejection, compared to a traditional LQ controller. By tailoring the sparse structure of the controller to the system dynamics, as well as the expected type of disturbances, and defining an optimization criterion that matches the desired controller properties, we have designed a sparse feedback controller that outperforms the classical LQ controller with regard to disturbance rejection. The sparse controller also preserves interpretability, making it more accessible for plant personnel to modify as the process changes over its life-cycle.

\bibliography{references}            

\end{document}